# Falling Bodies: the Obvious, the Subtle, and the Wrong[1]

Mario Rabinowitz
Electric Power Research Institute
Palo Alto, CA 94303

## Abstract

An important scientific debate took place regarding falling bodies hundreds of years ago, and it still warrants introspection. Galileo argued that in a vacuum all bodies fall at the same rate relative to the earth, independent of their mass. Aristotle seemed to consider all media to be viscous, and argued that heavier bodies fall faster. Aristotle was challenged by Philoponus, who argued that light and heavy weights fall about equally fast in air, eleven hundred years before Galileo. As we shall see, the problem is more subtle than meets the eye -- even in a frictionless medium. Philoponus and Galileo are right part of the time, and Aristotle is partly right some of the time. In fact they are all wrong the rest of the time, with the lightest body falling fastest when two bodies fall toward the earth. In principle the results of a free fall experiment depend on whether falling masses originate on earth, are extraterrestrial, are sequential or concurrent, or are simultaneous for coincident or separated bodies, etc. When single falling bodies originate from the earth, all bodies (light and heavy) fall at the same rate relative to the earth in agreement with Galileo's view. Einstein's General Relativity (EGR), in which gravity is due to space-time curvature, was motivated by the Galilean notion that free-fall is independent of the mass and properties of a falling body, and is just due to the properties of the milieu it finds itself in. Quantum mechanics is found to violate the Equivalence Principle of EGR.

## PHYSICS OVERVIEW

Despite the outstanding success of Einstein's General Relativity (EGR), there can be exceptions to the Equivalence Principle (EP) that is the very cornerstone upon which EGR is based. Due to increasing acceleration because of increased gravitational force as bodies fall toward each other, gravitational radiation with a concomitant retarding gravitational radiation reaction force can be expected. The EP is violated because the gravitational acceleration is independent of the accelerated mass, whereas the gravitational radiation reaction acceleration depends on the accelerated mass. This violation of the EP may only be perceptible for very high gravitational fields. There may also be quantum mechanical (QM) deviations from the EP. Particularly so since QM violates the EP.

[1]The PHYSICS OVERVIEW has been added and Sec. V has been expanded for clarification.

For historical context, only falling body effects will be focused on here. There are effects, such as the acceleration of the earth by the sun, which may be larger. As shown in this paper, the gedanken experiment with 3 gravitationally attracted masses, in which the lightest mass goes the fastest, is only a virtual violation of the EP.

## I. THE FALLING BODY PROBLEM

Many issues in science are never fully laid to rest. As well expected, there is generally disagreement when a subject is nascent. However, as humankind traverses the helix of knowledge and views the same subjects again and again from an ever higher vantage point with greater perspective and deeper insight, not only can disagreement recur but often a greater appreciation is instilled for what in the past had been discarded as obviously wrong. As we shall see, it is not always obvious what is right and what is wrong; and subtle distinctions are often better perceived in the hindsight of increased knowledge.

The falling body debate goes back to before 300 B.C. when Aristotle concluded that heavy bodies fall faster than light ones. In the 1600's Galileo set things straight by noting that heavy and light bodies fall at the same rate. In 1986, Fischbach et al reanalyzed the Eotvos experiment, noted anomalies in other data, and were led to conjecture a short-range fifth force that depending on composition could cause some light objects to fall faster than some heavy objects.[1]  Whether the fifth force holds or falls, their hypothesis will have served a useful purpose in plugging up a hole in a domain that had not been thoroughly studied.

I would like to use the falling body problem as an example of the subtleties that one can encounter in even what would appear to be a trivial problem. The magnitude of the effect that will be analyzed is extremely small because the earth is much more massive than most falling bodies.  Nevertheless, the spirit of the analysis is to illustrate that one may obtain seemingly wrong and contradictory experimental results even though they really are correct and compatible.  In the process we shall see that in some cases neither Aristotle nor Galileo were right, that a third outcome is even possible, and that Galileo was not first with his conclusions. In as unlikely a setting as the falling body problem we can also gain insight into a possible important additional reason why Einstein was on the negative side of the debate on the correctness of quantum mechanics.

## II.  ARISTOTLE, PHILOPONUS, and GALILEO

### II.1  Aristotle

In the 4th century B.C , Aristotle [2] reached the following conclusion about falling bodies: "If a certain weight moves [falls] a certain distance in a given time, a greater will move the same distance in a less time, and the proportion which the weights bear to one another, the times, too will bear to one another, e.g., if one weight is twice another, if the half weight cover the distance in x, the whole weight will cover it in x/2."

It is important to note that Aristotle is qualitatively correct in many real physical cases, and as we shall later see he is right in principle in some idealized cases. If we compare bodies of the same shape and size falling in a medium such as air or water,  then they do reach terminal velocities related to their weights as we will analyze in a later section.  Of course the terminal velocity is reached much sooner in the denser medium.  If Aristotle did conduct experiments, he likely did it in a liquid such as water to slow down the falling body to a more easily observable speed than in air.

It is difficult to judge whether Aristotle conducted falling body experiments. In cases where a simple observation could have avoided him an obvious error, he clearly did not conduct experiments. He sometimes made astute empirical observations as when he argued for a spherical earth not only because of the "perfect" shape of a sphere, but also because the earth casts a circular shadow on the moon during a lunar eclipse.  However, he did not make the connection between free fall and the inclined plane that enabled Galileo to reach and easily verify his conclusions.

### II.2 Philoponus

From an historical point of view, it is not true that Galileo was the first to challenge Aristotle and in so doing to introduce the experimental method. As early as the 6th century A.D., Ioannes Philoponus [3] challenged Aristotle: "But this is completely erroneous, and our view may be corroborated by actual observation more effectively than by any sort of verbal argument.  For if you let fall from the same height two weights of which one is many times as heavy as the other, you will see that the ratio of the times required for the motion does not depend on the ratio of the weights, but that the difference in time is a very small one."

**II.3 Galileo**

Eleven hundred years later, in the early 17th century, Galileo [4] made essentially the same observation as Philoponus: "But I, Simplicia, who have made the test, can assure you that a cannon ball weighing one or two hundred pounds, or even more, will not reach the ground by as much as a span ahead of a musket ball weighing only half a pound, provided both are dropped from a height of 200 cubits."

He argued that the slight difference in time could be ascribed to the resistance offered by the medium to the motion of the falling body. In air, feathers do fall more slowly than rocks. Galileo then made the idealization that in a medium devoid of resistance (a vacuum), all bodies will fall at the same speed. This idealization neglected the complexity of the fall of objects in media accessible to Galileo and his predecessors, and was indeed a significant advance toward a deeper understanding of the motion of bodies.

**III. MOTION OF GRAVITATIONALLY ATTRACTED BODIES**

Consider two masses m and M interacting with one another gravitationally in vacuum. The force of mutual attraction is $F = -GmM/r^2$, where G is the universal gravitational constant and r is the distance between the centers of m and M. [We will assume spherical symmetry for the bodies.] Since this is a central force, the motion of the two bodies about their center of mass (CM) can be formally reduced to an equivalent one-body problem with a body of reduced mass $\mu = mM/(m + M)$. The motion of m as viewed from M is the same as if M were fixed, and m had the reduced mass µ.

If the bodies are released from rest with initial separation r and final separation R, by the principle of conservation of energy we have

$$\tfrac{1}{2}mv^2 + \frac{-GMm}{R} = 0 + \frac{-GMm}{r}, \tag{1}$$

where v is the relative velocity acquired by the two bodies as they fall toward one another. Solving Eq. (1)

$$v = \left[2G(M+m)(R^{-1} - r^{-1})\right]^{1/2}. \tag{2}$$

From Eq. (2) we see that the velocity of approach of the two masses is proportional to the sum of the two masses, as is the acceleration

$$a = dv/dt = \frac{G(M+m)}{r^2}. \tag{3}$$

The above derivation is for measurements made relative to M (the earth). [We will neglect acceleration of the earth by the sun, Coriolis, centripetal, tidal force, relativistic, etc. effects in our idealized experiments.] We see from Eqs. (2) and (3) that when M > > m, v and a are to an excellent approximation independent of m and just $\propto$ M. However, as we shall see, when we are interested in precise results it makes an important difference if the measurements are relative to one of the bodies, or the center-of-mass of the system (CM). It further makes a difference if m + M is held constant or not. If the body originates from the earth we have (M – m) + m = M. So all terrestrial bodies falling singly -- regardless of the size of m -- fall at exactly the same rate as long as removal of m from the earth does not perturb the earth's approximate spherical symmetry.

**IV. GEDANKEN EXPERIMENTS**

The following gedanken (thought) experiments have virtual anomalies, i.e., they appear to be contradictory, but are really compatible. All the bodies will be dropped from the same height in vacuum, and have the same size to avoid problems related to differences in radius in contacting the earth. Let us first look at experiments where the bodies (masses) originate from the earth, and then contrast this with experiments where the masses have an extraterrestrial origin. In a third set of experiments, the origin of the bodies is not important. In all the experiments, it will make a difference (but in variable ways) depending on whether the measurements are made relative to the earth or the CM. The experimental results appear to conflict, and it appears that some must be wrong. However, we will see that the results can be reconciled and that the most surprising conclusion is that they are all correct.

One remarkable conclusion is that Galileo, Philoponus, and Aristotle are strictly incorrect for a set of experiments in which light bodies fall faster than heavy ones. Galileo is rigorously and exactly correct for one set of experiments in which all singly falling terrestrial bodies fall at the same rate. Aristotle is correct (heavy bodies fall faster than light ones) for another set. It's an illustration that sometimes one can obtain comprehensible, correct but otherwise seemingly disparate experimental results.

**IV.1 Earth Masses**

Let us take two masses ($m_2 > m_1$) originating from the earth of mass $M_e$. We drop $m_1$ and it falls to the earth in time $t_1$. In a similar experiment, $m_2$ reaches the earth in time $t_2$. We find that the two bodies hit the earth in exactly equal time, $t_1 = t_2$. In accord with Galileo, we find in this experiment that **all earth masses dropped sequentially fall at exactly the same rate relative to the earth** since $(M_e - m_1) + m_1 = (M_e - m_2) + m_2 = M_e$, as given by Eq. (3).

The case of simultaneous free fall of two bodies toward a third body (e.g. the earth) is more subtle, and is treated in Sections V and VI. As we shall see, there is a case relative to the CM, where neither Aristotle, Philoponus, nor Galileo are correct.

**IV.2 Extraterrestrial Masses**

Now let us bring a mass $M_1$ ($M_1 = m_1$) in from outer space. When dropped it reaches the ground in time $T_1$. We return $M_1$ to outer space, and come back with a mass $M_2$ ($M_2 = m_2 > M_1$). Mass $M_2$ has a descent time $T_2$. We find that $T_2 < T_1$. In accord with Aristotle, we find here that heavy bodies fall faster than light bodies relative to the earth. [However as we shall see, relative to the CM Galileo is right.] This is a real contradiction to Philoponus and Galileo, and seems contradictory to the earth masses experiment, but isn't as we shall see. To compound the apparent inconsistency, $t_1 = t_2 > T_1 > T_2$. Had we left mass $M_1$ on the earth, when we went back to obtain $M_2$, in our experiment with $M_2$ we would have measured a fall time $T_3 < T_2$.

**IV.3 Resolution of the Results**

The results of these two sets of experiments are consistent with each other and in accord with the analysis derived in this paper. The velocity of the dropped mass relative to the earth is proportional to the sum of the masses as given by Eq. (2). In the first set of experiments, the sum of all the masses remains constant, because the test body (heavy or light) always originates from the earth. Therefore the test body's velocity and fall time is independent of its mass, not just because its mass factors away (as taught in textbooks), but because its mass plus the mass of the remainder of the earth is constant:

$v \propto m + (M_e - m) = M_e$ (the earth's mass).

Equivalently, the net acceleration a with respect to the earth is the vector difference of the individual accelerations with respect to the system center of mass (CM). Thus

$$a = a_M - a_m = \frac{Gm}{r^2} - \frac{-GM}{r^2} = \frac{G(m+M)}{r^2}. \qquad (5)$$

This is just another way to obtain Eq. (3). In the first set of experiments, a is independent of m since $m + (M_e - m) = M_e$ = constant. Curiously, this very fact leads to the conclusion that **for terrestrial bodies relative to the CM, the lighter the body the faster it falls** because $a_m = GM/R^2 = G[M_e - m_1]/R^2 > G[M_e - m_2]/R^2$.

In the second set of experiments **with extraterrestrial bodies**, the test mass is added to the mass of the earth ($M_e$) in Eq. (2). Thus $v \propto M + M_e > M_e$. This is why the time of descent relative to the earth progressively decreases ($T_1 > T_2 > T_3$) as the sum of the masses taking part in the experiment progressively increases. **So relative to the earth experiments with extraterrestrial bodies favors Aristotle. For extraterrestrial bodies, relative to the CM the body's mass factors out and $a_M = GM_e/R^2$ making Philoponus and Galileo right.**

So we have to be careful as to the source of the mass, and whether the measurements are relative to the earth or the center-of-mass of the system (CM).

## V. SIMULTANEOUS FREE FALL OF 3 SEPARATED BODIES

One may think that the results of the previous gedanken experiments are an artifact resulting from the sequential nature of the tests. Surely Aristotle, Philoponus, and Galileo were contemplating simultaneous free fall experiments of the bodies relative to the earth since clocks were not that accurate in those days. Which view is correct for simultaneous free fall?

**For a given simultaneous free fall experiment, it does not make any difference whether the bodies are terrestrial or extraterrestrial in origin.** The three-body problem is more difficult to analyze. Let us free ourselves from any unnecessary complexities by considering three bodies free from any other interactions except their central force interaction. The bodies could be spheres of masses $M >> m_2 >> m_1$; or even idealized point masses.

In the first thought experiment of the next series, let the three masses be in a straight line with M an equal distance between $m_2$ and $m_1$. (This would be like dropping $m_2$ and $m_1$ simultaneously from equal heights at opposite poles of a spherical earth.) When let go, the three bodies accelerate toward their common center of mass. Since the center of mass of the system is between the centers of M and $m_2$, $m_2$ will have a shorter distance to fall toward the CM. Since M moves towards $m_2$, M moves away from $m_1$. Thus $m_1$ will have a longer distance to fall than $m_2$ to reach M or the CM. Therefore relative to M (the earth), one might be tempted to think that the heavier mass $m_2$ will fall faster than $m_1$. However, all three bodies must reach the CM at the same instant or the CM would move, which is not allowed. **Since the lightest mass $m_1$ has to go the farthest distance to reach the CM, it must go the fastest relative to the CM.**

**Let us analyze the 3-body rectilinear gravitational force problem in the CM system**, where the bodies are placed collinearly on the x-axis in the sequence $m_1$, M, and $m_2$ ; and their coordinates have the sequence $x_1$, $x_M$, and $x_2$. At time t = 0:

$$m_1 a_1 = \frac{GMm_1}{(x_M - x_1)^2} + \frac{Gm_2 m_1}{(x_2 - x_1)^2} \Rightarrow a_1 = \frac{GM(x_2 - x_1)^2 + Gm_2(x_M - x_1)^2}{(x_M - x_1)^2 (x_2 - x_1)^2}. \tag{6a}$$

Similarly

$$a_2 = \frac{-GM(x_2 - x_1)^2 - Gm_1(x_2 - x_M)^2}{(x_2 - x_M)^2 (x_2 - x_1)^2} = \frac{-GM(x_2 - x_1)^2 - Gm_1(x_M - x_1)^2}{(x_M - x_1)^2 (x_2 - x_1)^2}, \tag{6b}$$

since $(x_M - x_1) = (x_2 - x_M)$ because the masses $m_1$ and $m_2$ are initially placed equidistant from M. Dividing Eq. (6a) by (6b) we obtain the magnitude of the ratio of the initial accelerations of $m_1$ and $m_2$ relative to the CM:

$$\left|\frac{a_1}{a_2}\right| = \frac{M(x_2 - x_1)^2 + m_2(x_M - x_1)^2}{M(x_2 - x_1)^2 + m_1(x_M - x_1)^2} = \frac{4M + m_2}{4M + m_1} > 1. \tag{7}$$

Because $m_2 > m_1$ , and $(x_2 - x_1) = 2(x_M - x_1)$ initially.

Starting with Eqs. (6), the analysis for the 3-body rectilinear motion problem utilized Newton's gravitation law explicitly in obtaining Eq. (7). **Let us now see what we can learn without explicit knowledge of the interaction forces between the 3 bodies, purely from the fact that there is no force acting on the center of mass (CM) of the 3-body system.** As before, the 3 bodies are placed collinearly on the x-axis in the sequence $m_1$, M, and $m_2$. The coordinates have the sequence $x_1$, $x_M$, $x_{CM}$, and $x_2$. The masses $m_1$ and $m_2$ are initially placed equidistant from M so that $(x_M - x_1) = (x_2 - x_M)$, and then allowed to move toward each other under the action of mutually attractive forces. Since there is no external force, as the 3 bodies move together, the center of mass of the system which is an inertial frame (either at rest or moving uniformly) must remain fixed with respect to our coordinate system in the CM frame.

For convenience we set the center of mass coordinate $x_{CM} = 0$. At t = 0:

$$x_{CM} = 0 = \frac{m_1 x_1 + M x_M + m_2 x_2}{m_1 + M + m_2} \Rightarrow x_1 = \frac{(M + 2m_2) x_M}{m_1 - m_2} \Rightarrow a_1 = \frac{d^2 x_1}{dt^2} = \frac{(M + 2m_2)}{m_1 - m_2} \left( \frac{d^2 x_M}{dt^2} \right). \quad (8a)$$

Similarly

$$a_2 = \frac{d^2 x_2}{dt^2} = \frac{(M + 2m_2)}{m_1 - m_2} \left( \frac{d^2 x_M}{dt^2} \right). \quad (8b)$$

Dividing Eq. (8a) by Eq. (8b), so at t = 0:

$$\left| \frac{a_1}{a_2} \right| = \frac{M + 2m_2}{M + 2m_1} > 1. \quad (9)$$

It is noteworthy (if not strange) that Eqs. (7) and (9) are significantly different for 3 bodies. Yet, in the case of 2 bodies, if we set M = 0, they both yield the same result $\left| \frac{a_1}{a_2} \right| = \frac{m_2}{m_1}$, where we have obtained Newton's 3rd Law for this simple case.

**So we find the intriguing result that relative to a point fixed in space (the CM) for 3 bodies, the lighter body falls faster that the heavy body for bodies falling concurrently from opposite ends of the earth, making Philoponus, Galileo, and Aristotle all wrong in this case. The same conclusion that $m_1$ is faster than $m_2$ applies even if $m_1$, $m_2$, and M are not collinear**, as we shall see next.

Now consider the masses $m_2$ and $m_1$ brought closer together so that they form an angle between 0º and 180º relative to the center of mass. Again the center of mass of the system is closer to $m_2$ than $m_1$. Hence, as the three bodies fall toward their common center of mass (which must remain at rest since there are no external forces}, $m_1$ has a further distance to reach the CM than $m_2$. Hence $m_1$ the lighter body will always fall faster than $m_2$ relative to the CM for any angle of separation >0º. Again $m_2$ and $m_1$ will hit the earth in equal times, given that they have the same radius.

For other than 180° separation, they will not fall in straight lines as they also are attracted to one another as well as moving toward the common center-of-mass of the system. (Equal masses at the vertices of an equilateral triangle would move in straight lines.) The lightest body is further handicapped because its trajectory deviates the most from a straight line. The curved trajectories will maintain 0 total angular momentum. As $m_2$ and $m_1$ get closer and closer together, their mutual attraction will dominate over their free fall toward M. To avoid this, their initial separation should be large enough that they don't collide before they hit the earth.

**VI. SIMULTANEOUS FREE FALL OF COINCIDENT BODIES**

When the concurrently falling bodies were separated, the lighter mass fell faster than the heavier mass relative to the CM, and the same rate relative to the earth. For our next thought experiment, we want to see what happens when we drop $m_2$ and $m_1$ from the same point simultaneously. This 0° case has to be done carefully to avoid the criticism, "of course they fell at the same rate since they were stuck together gravitationally and acted as one body." Let $m_2$ be a large hollow transparent sphere with $m_1$ a tiny sphere inside it at its center. (The inner and outer spheres have no net gravitational attraction between them for all points inside the outer sphere.) The center of mass of the system is in line with $m_2$ and $m_1$, and equally distant from both, as is M. **Therefore they will fall at the same rate with respect to both the earth and the center of mass of the system.** ($m_2$ and $m_1$ could be reversed for further verification.)

**VII. EQUIVALENCE PRINCIPLE (EP)**

In the previous 3-body analyses, it appears as if the principle of the equivalence of inertial and gravitational mass in Einstein's General Relativity is being violated because the lightest mass has the highest acceleration. As shown this is not a violation of the EP since the inertial and gravitational masses are equal. The lightest mass $m_1$ goes the fastest in the center of mass 3-body system because it is acted by both M and $m_2$, whereas $m_2$ is acted on by both M and $m_1$.

**VIII. GALILEO'S ARGUMENT**

Galileo used experiments with an inclined plane to promote his view that heavy and light bodies fall equally fast. However to show that Aristotle's hypothesis is logically inconsistent, he felt it necessary to present a rhetorical argument [5]: "Tie m, a light stone, together with M, a heavy one, to form a double stone. Then in falling, m should retard M, since it falls more slowly than M. Hence the combination should fall at some speed between that of m and M. However. according to Aristotle, the double body (m + M), being heavier than M, should fall faster than M."

Galileo presents this reductio ad absurdum argument not only to show the fallacy of Aristotle's logic, but since the body (m + M) cannot fall both more slow!y and more quickly than the body M, it must therefore fall at the same speed as M. However, the analysis in this paper shows that even in vacuum for bodies with an extraterrestrial origin dropped sequentially, the double body (m + M) does indeed fall faster than M.

Galileo's logic is non-sequitur. To see the fallacy in Galileo's logic, let us consider a simple example. Let us drop two hollow bodies of the same size, but of different densities, in a viscous fluid such as water or oil. The heavy one, M, will fall faster than the light one, m. Next compact one so that it fits inside the other. When the double body (m + M) falls in the fluid. it will fall faster than M alone. If the bodies are joined externally, the lighter body may slow down the heavy one, but it is a different problem related to area rather than mass alone.

Galileo, Philoponus, and Aristotle did not distinguish between free fall relative to the earth and relative to the center-of-mass system, as the CM concept came later. Interestingly, physics texts, and philosophy of science texts in presenting Galileo's demolition of Aristotle's premise, are not concerned with the origin of the falling mass, and the distinction between relative to the earth and the CM. This is probably because the $6 \times 10^{24}$ kgm mass of the earth is so great compared to that of any test body that the distinction hardly seems worth the bother.

## IX. VISCOUS MEDIA AND ARISTOTLE

When the bodies fall through a viscous medium, the viscosity depends on the body's velocity. When only a small velocity is acquired, corresponding somewhat to laminar flow around light bodies, the viscous drag force $\propto$ the body's velocity in close accord with Aristotle. When a large velocity is acquired, corresponding somewhat to turbulent flow around heavy bodies, the drag force $\propto$ the body's (velocity)$^2$.

## X. EINSTEIN, RELATIVITY, AND QUANTUM THEORY

Taking the falling body problem into the domain of quantum mechanics leads to a remarkable result. Let us first note that gravitational satellite (or planetary) motion is akin to the falling body problem. A satellite falls in toward the earth just like a dropped body, except that it has a high enough tangential velocity that it keeps missing the earth as it falls. If the rate of fall of a body (relative to an inertial frame) is independent of its mass, then this indicates an equivalence of inertial and gravitational mass. The equivalence principle (EP) in a more general form is the basis of gravitation in Einstein's theory of General Relativity. Einstein concluded that since the gravitational acceleration of a freely falling mass does not appear to depend upon any of the properties of the body, it may be considered to be a property of the geometry of space-time. He thus postulated an equivalence locally between a gravitational field and an accelerating frame.

In classical (non-quantum) mechanics with respect to the CM, the orbital radius, r, of a mass, m, held in orbit by a mass M is: r = GM/v², independent of m. However, quantum mechanically (simple Bohr theory here, but all of QM violates the EP) the allowed radius is

$$r = \frac{n^2 h^2}{GMm^2(2\pi)^2}, \tag{12}$$

where $\hbar$ is Planck's constant/$2\pi$, and n is an integer. The allowed Bohr quantum gravitational acceleration is:

$$a = \frac{GM}{r^2} = \frac{G^3 M^3 m^4 (2\pi)^4}{n^4 h^4}. \tag{13}$$

This is at odds with the equivalence principle, and may possibly be a reason in addition to the probabilistic aspect of quantum theory that Einstein was uncomfortable with quantum mechanics. The quantum gravitational radius and acceleration are not independent of m. This may be a critical stumbling block to a quantized theory of general relativity. The equivalence principle is a concept that applies loca1ly. In quantum mechanics expectation or mean values (obtained by integrating the wave function over all space) correspond to measured values. A superb neutron interferometer experiment in which the neutrons were quantum mechanically affected by gravity was conducted by Colella, Overhauser, and Werner [5]. It sheds light on this important question. This experiment is delightfully described by Greenberger and Overhauser [6]. Interestingly, if the quantization condition for angular momentum in a gravitational field were different than the usual $L = n\hbar$, then r and a could be independent of the falling mass, m.

**XI. CONCLUSION**

In free fall between bodies of comparable mass, one must apply precise analysis because the standard approximation would fail in many cases. As we've seen, one can sometimes look at something long taken for granted, and if one is patient enough one can uncover very interesting subtleties. There are both absolute and relative aspects to the findings just as the sequence of colors in a rainbow will always be the same to all observers, and yet the rainbow itself is a function of the observer's position -- be it an eye or a camera. When a ball is dropped from a bridge out of a car moving at constant velocity, its acceleration is the same with respect to the car or an observer at rest. Yet the ball's trajectory is straight down relative to the car, and is a parabola relative to a stationary observer. So, it shouldn't be too perplexing if we find that all three possibilities: heavy faster than light, heavy and light equally fast, and even light bodies faster than heavy bodies can all occur in the falling body problem.

**Acknowledgment**

I wish to thank Felipe G. Garcia, Arthur Cohn, and David Frankel for stimulating and helpful discussions. Special thanks are due to Michael Ibison for an enjoyable and helpful dialogue on the recently added material in Sec. V.